\pdfoutput=1 

\documentclass[%
,aps%
 ,preprint,tightenlines%
 ,secnumarabic%
 ,nofootinbib
,amssymb, amsmath,nobibnotes, prl, floatfix]{revtex4}
\usepackage{docs}%
\usepackage{bm}%
\expandafter\ifx\csname package@font\endcsname\relax\else
 \expandafter\expandafter
 \expandafter\usepackage
 \expandafter\expandafter
 \expandafter{\csname package@font\endcsname}%
\fi

\usepackage[italian, english]{babel}
\usepackage{amsmath}                                      
\usepackage{amsthm}

\usepackage{cancel}

\usepackage{wrapfig}
\usepackage{amssymb}
\usepackage{comment}
\usepackage{latexsym}
\usepackage{mathtools}
\usepackage{dsfont}
\usepackage{mathrsfs}
\usepackage{hyperref}
\usepackage{float}
\usepackage{graphicx}

\usepackage{framed}
\usepackage{amsfonts}

\usepackage{pgf}
\usepackage{physics}

\newcommand{\1}{\mathds{1}}
\newcommand{\be}{\begin{equation}}
\newcommand{\ee}{\end{equation}}

\newcommand{\beq}{\begin{equation}}

\newcommand{\eeq}{\end{equation}}

\newcommand{\ch}{\cosh}

\newcommand{\bs}{\begin{split}}
\newcommand{\es}{\end{split}}

\newcommand{\pd}{\partial}

\newcommand{\vol}{\mathrm{vol}}

\newcommand{\ads}{\ensuremath{\operatorname{AdS}}}
\newcommand{\ds}{\ensuremath{\operatorname{dS}}}
\newcommand{\cft}{\ensuremath{\operatorname{CFT}}}
\newcommand{\sent}{\ensuremath{S_\text{ent}}}
\newcommand{\sltr}{\ensuremath{SL(2,\mathbb{R})}}

\newcommand{\Leff}{\ensuremath{L_\text{eff}}}

\numberwithin{equation}{section}
\allowdisplaybreaks[4]

\definecolor{cardinal}{rgb}{0.6,0,0}
\definecolor{darkgreen}{rgb}{0,0.4,0}
\definecolor{amethyst}{rgb}{0.54, 0.17, 0.89}
\definecolor{cerulean}{rgb}{0.0, 0.48, 0.65}

\begin{document}


\makeatletter
\preto\maketitle{%
  \begingroup\lccode`~=`,
  \lowercase{\endgroup
  \let\saved@breqn@active@comma~
  \let~}\active@comma 
}
\appto\maketitle{%
  \begingroup\lccode`~=`,
  \lowercase{\endgroup
  \let~}\saved@breqn@active@comma 
}
\makeatother

\title{AdS Vacuum Bubbles, Holography and Dual RG Flows}

\author{Riccardo Antonelli}
\email{riccardo.antonelli@sns.it}
\author{Ivano Basile}
\email{ivano.basile@sns.it}
\affiliation{{Scuola Normale Superiore and I.N.F.N.\\Piazza dei Cavalieri 7, 56126, Pisa, Italy}}

\author{Alessandro Bombini}
\email{alessandro.bombini@pd.infn.it}
\affiliation{Department of Physics and Astronomy ``Galileo Galilei'' \\ I.N.F.N. Sezione di Padova\\ via Marzolo 8, 35131, Padova, Italy}

\date{\today}%

\begin{abstract}
We explore the holographic properties of non-perturbative vacuum decay in Anti-de Sitter ($\ads$) geometries. To this end, we consider a gravitational theory in a metastable $\ads_3$ state, which decays into an $\ads_3$ of lower vacuum energy via bubble nucleation, and we employ the Ryu-Takayanagi conjecture to compute the entanglement entropy $\sent$ in its alleged holographic dual. Our analysis connects the nucleation and growth of a vacuum bubble to a relevant deformation and a subsequent Renormalization Group (RG) flow in the boundary theory, with $\sent$ a $c$-function. We provide some evidence for the claim and comment on the holographic interpretation of off-centred or multiple bubbles. We also frame the issue in the formalism of Holographic Integral Geometry, highlighting some consequences on the structure of the holographic RG flow and recovering the standard holographic RG as a limiting case. 
\end{abstract}

\maketitle
\tableofcontents



\section{Introduction} \label{sec:intro}

Despite significant advances and crucial insights obtained in decades of research, Quantum Gravity remains a remarkably difficult challenge. The main available tools stem from dualities, which are best understood in supersymmetric scenarios, and from holography in geometries of the Anti-de Sitter ($\ads$) type~\cite{Maldacena:1997re, Witten:1998qj, Gubser:1998bc}.  
The semi-classical limit of Quantum Gravity appears more manageable and universal in an effective framework. In particular, black-hole thermodynamics has proven a fruitful angle of attack for these issues~\cite{Bardeen:1973gs}. Moreover, black holes are also well understood holographically~\cite{Strominger:1996sh, Strominger:1997eq, David:2002wn}, at least in regimes in which the gravitational side is under control. The holographic properties of black holes are encoded in thermal states of the corresponding boundary theories, and (entanglement) entropy computations provide a useful tool to study them~\cite{Ryu:2006bv, Rangamani:2016dms, Hubeny:2007xt}.

    All in all, black holes constitute a prototypical example of a quantum gravitational phenomenon. Similarly, vacuum decay processes~\cite{Coleman:1980aw, Brown:1987dd, Brown:1988kg} comprise a different class of scenarios where genuine quantum gravitational effects drive the physics. Much as for black holes, the semi-classical description of vacuum decay has been thoroughly dissected in the literature~\cite{Kanno:2011vm, Freivogel:2016qwc, Ooguri:2016pdq, Danielsson:2016rmq} and is currently an active topic of research, but its holographic properties were only explored to a lesser extent\footnote{For recent results, which appeared during the development of this project, see~\cite{Burda:2018rpb, Hirano:2018cyr}. See also~\cite{deHaro:2006ymc, Papadimitriou:2007sj} for other works on the structure of the vacuum in the presence of bubbles. For a field-theoretical discussion of instanton contributions to entanglement entropy, see~\cite{Bhattacharyya:2017pqq}.}. The issue was investigated in connection with the walls of vacuum bubbles~\cite{Maldacena:2010un, Barbon:2010gn, Harlow:2010az}, but here we would like to explore the links with the boundary of $\ads$, which suggests a qualitatively different picture.
Since vacuum decay processes also play an important role in identifying a ``swampland'' and its relation to UV completions of gravity, it is conceivable that probing them beyond the semi-classical level could provide new gateways to the intricacies of the field~\cite{Brennan:2017rbf}.

In this paper we propose a first step to bridge the gap between holographic methods, which typically address stable, often exclusively stationary states, and aspects of the standard semi-classical techniques used to study vacuum decay, focusing in particular on the development of vacuum bubbles that mediate transitions between classical vacua.  Here we consider them in the simplest case of interest, $\ads$ geometries in $D=3$ dimensions.
We find evidence that, holographically, vacuum bubbles behave much like renormalisation group flows of the boundary theory, and appear to provide, in some sense, a set of building blocks for such flows, as we shall discuss later on. The motivation for considering this interpretation relies on two facts:
\begin{itemize}
\item Vacuum decay has an irreversible  direction, from $\ads$ radius $L_-$ to $L_+ < L_-$, \textit{i.e.} the (negative) cosmological constant must increase in absolute value~\cite{Brown:1987dd,Brown:1988kg}. 
\item The central charge, in an $\ads_3$ vacuum, is proportional to the $\ads_3$ radius, in particular
\begin{equation}\label{eq:c_dictionary}
   c = \frac{3 L}{2 G_3} 
\end{equation}
in three dimensions~\cite{Brown:1986nw}, where $G_3$ is the three-dimensional Newton constant. This suggests that vacuum decay be accompanied by a \textit{decrease} of the central charge $c$, along the lines of the Zamolodchikov $c$-theorem~\cite{Zamolodchikov:1986gt}. Our choice of working in $D=3$ is indeed motivated by the fact that, while gravity becomes more tractable~\cite{Witten:1988hc, Carlip:1991ij}, the central charge encodes key information on the boundary theory~\cite{Cardy:1986ie, Calabrese:2004eu, Calabrese:2009qy}.
\end{itemize}
In order to put this idea on firmer grounds, it will be useful to study the behaviour of the entanglement entropy (EE) of any subregion of the deformed boundary theory, since this quantity provides a probe for its quantum behaviour.
If this framework gives a correct description of the problem, important lessons are potentially in store regarding the swampland program and the stability of non-supersymmetric Anti-de Sitter ``vacua''. Moreover, powerful standard techniques that apply to the boundary description could conceivably shed light on the analysis of vacuum instabilities beyond the semi-classical regime.


\section{Bubble Growth in $\ads$}

\label{sec:bubble_geometry}

In this section we present the geometry which models the decay process that we shall consider. It describes, in the semi-classical limit, the expansion of a bubble of $\ads$ geometry, nucleated by tunneling, inside a metastable $\ads$ of higher vacuum energy. Physically, such a situation can be realised, in the simplest setting, in a gravitational theory with a minimally coupled scalar subject to an asymmetric double well potential~\cite{Coleman:1980aw, Freivogel:2016qwc} of the form
\begin{equation}\label{eq:proto_lagrangian}
    \mathcal{L} = R -\frac{1}{2}(\partial \Phi)^2 - V_{\text{well}}(\Phi) \, .
\end{equation}

In the following we shall not need a precise form for the potential $V_{\text{well}}$, since in this paper we focus on model-independent features. Nevertheless, we remark that a more explicit top-down construction, possibly in terms of non-supersymmetric string models~\cite{AlvarezGaume:1986jb, Sagnotti:1995ga, Sagnotti:1996qj, Sugimoto:1999tx, Antoniadis:1999xk, Angelantonj:1999jh, Aldazabal:1999jr, Angelantonj:1999ms, Dudas:2000nv, Pradisi:2001yv, Angelantonj:2002ct, Mourad:2017rrl} or orbifold field theories~\cite{Kachru:1998ys,Lawrence:1998ja,Bershadsky:1998mb,Bershadsky:1998cb,Schmaltz:1998bg,Erlich:1998gb}, should provide better control of the holographic dictionary in this context. We intend to address this issue in a future work.

In order to isolate the relevant physics in the most tractable scenario, we shall work in $D=3$, while resorting to the thin-wall approximation. Furthermore, we shall focus on nucleation at vanishing initial radius\footnote{Since bubble nucleation is a genuinely quantum-gravitational event, one may expect tunneling to favor Planck-scale initial radii. Therefore, at the semi-classical level we expect our approximation to be instructive.}, occurring at the centre of a global chart of an original $\ads_3^-$ spacetime. The generalization to arbitrary initial radius is straightforward and does not appear to affect our analysis qualitatively, while off-centre nucleation is discussed later.

\subsection{Construction of the Geometry}

Let us consider two $\ads_3$ vacua, dubbed $\ads_3^-$ and $\ads_3^+$, of radii $L_- > L_+$ respectively, connected by a tunneling process
\begin{equation}\label{eq:process}
\ads_3^- \longrightarrow \ads_3^+
\end{equation}
mediated by the nucleation of a bubble. Working in the thin-wall approximation, we realise the metric corresponding to the decay process by gluing the two $\ads_3$ vacua over a null surface, which represents the bubble trajectory.

\begin{figure}
    \centering
    \includegraphics[width=3.5in]{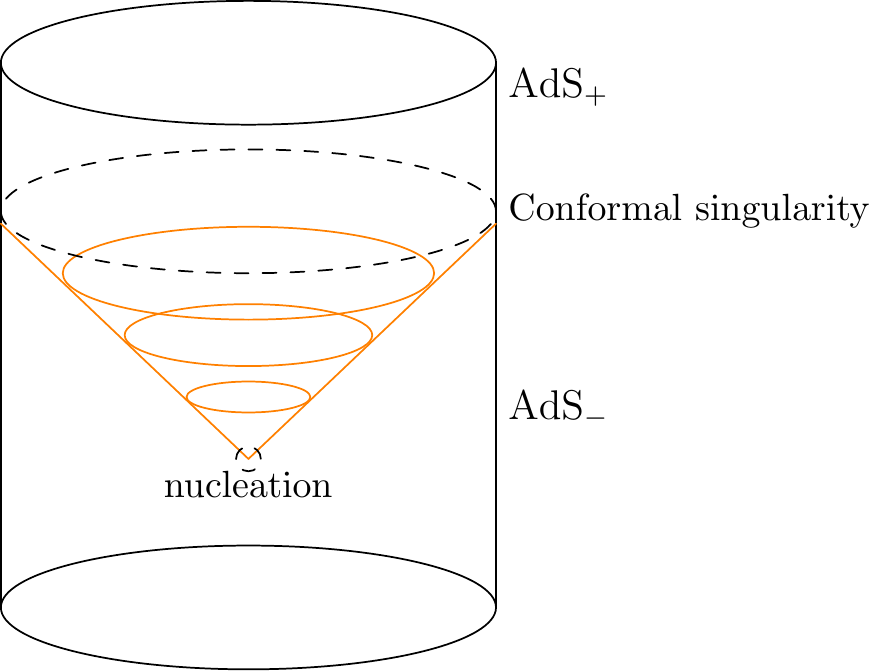}
    \caption{a Penrose-like diagram of the geometry describing the decay process.}
    \label{fig:penrose}
\end{figure}

It is most convenient to work in the following chart\footnote{This is related to the $(t,r,\phi)$ global coordinates via the transformation $\eta = L_\pm \tan(t/L_\pm)$. This chart does not cover the full geometry, but it does cover the entirety of the collapse.} for both the initial and final $\ads_3$ geometries,
\begin{equation}
    ds^2_\pm = - \left(1+\frac{r^2}{L_\pm^2}\right) \frac{d\eta^2}{\big(1+\frac{\eta^2}{L_\pm^2}\big)^2} + \frac{dr^2}{1+ \frac{r^2}{L^2_\pm}} + r^2 d\phi^2\, .
\end{equation}

In the thin-wall approximation the bubble is described in the $\ads^\pm_3$ charts respectively by the radial null surfaces
%
\begin{equation}\label{eq:bubble_motion}
ds^2_\pm = 0  \quad  \Longrightarrow \quad r = \eta\, .
\end{equation}
Gluing along the bubble\footnote{We remark that this can be done without problems, since the bubble is null. Equivalently, the equations for the bubble trajectory, seen from both sides, take the same form, which motivates this choice of time coordinate.}, the complete metric can be written in the compact form
\begin{equation}\label{eq:bubble_metric}
ds^2 = -\bigg(1 + \frac{r^2}{L^2_{\text{eff}}} \bigg)\frac{d\eta^2}{\big(1+\frac{\eta^2}{L^2_{\text{eff}}}\big)^2} + \frac{dr^2}{1 + \frac{r^2}{ L^2_{\text{eff}}}} + r^2 d\phi^2 \, ,
\end{equation}
where $L_{\text{eff}}$ denotes an \textit{effective curvature radius}, defined by
\begin{equation}\label{eq:eff_L}
L_{\text{eff}}(\eta,r) \equiv \begin{cases}
L_-\, , & r < \eta\\
L_+\, , & r > \eta
\end{cases} \, .
\end{equation}
It should be noted that $L_{\text{eff}}$ can be written as a step function with argument $r - \eta$. This may lead one to expect that doing away with the thin-wall approximation could amount to a smoothing of $L_{\text{eff}}$, perhaps as a function of an invariant quantity, which we shall indeed identify in the next section.

This gluing procedure agrees with the standard Israel junction conditions for null hypersurfaces~\cite{Israel1966, Barrabes:1991ng,Poisson:2002nv}. Indeed, the continuity condition for the (degenerate) induced metric $h$ on the bubble reduces to eq.~\eqref{eq:gluing}, while the transverse curvature experiences a discontinuity proportional to $h$, which can be ascribed to the bubble energy-momentum tensor~\cite{Poisson:2002nv}. In detail, following the notation of~\cite{Poisson:2002nv}, in the global $(\eta , r, \phi)$ chart the bubble (where $\eta = r$) is described by $\phi$, generated by the integral flow of the tangent space-like vector $e_\phi$, and by the null coordinate $\lambda \equiv \eta + r$, generated by the integral flow of the null vector $e_\lambda$. In addition, the transverse null vector $N$ is chosen such that
\begin{equation}\label{eq:tangent_normal_conditions}
    e_\phi \cdot e_\lambda = e_\phi \cdot N = 0 \,, \quad N^2 = e_\lambda^2=0 \,, \quad N\cdot e_\lambda = -1 \,.
\end{equation}
Explicitly,
\begin{equation}\label{eq:tangent_vectors}
e_{\lambda} \equiv \sqrt{\frac{f_\pm(r)}{2}} \, (\partial_\eta + \partial_r) \, , \qquad e_{\phi} \equiv \frac{1}{r} \, \partial_\phi \, , \qquad N \equiv \sqrt{\frac{f_\pm(r)}{2}} \, (\partial_\eta - \partial_r)
\end{equation}
on either side of the bubble, where $f_\pm(r) \equiv 1 + r^2/L^2_\pm$. The resulting transverse curvature
\begin{equation}\label{eq:transverse_curvature}
C_{ab} \equiv - \, g_{\mu \nu} \, N^\mu \, e^\rho_a \, \nabla_\rho \, e^\nu_b \,, \quad a,b \in \{ \lambda, \phi\} 
\end{equation}
is then
\begin{equation}\label{eq:transverse_curvature_components}
    C_{\lambda \lambda} = C_{\lambda \phi} = 0\,,\quad C_{\phi \phi} = \frac{1}{r} \, \sqrt{\frac{f_\pm(r)}{2}} \, .
\end{equation}
Hence, $C_{ab}$ is proportional to the (degenerate) induced metric $h_{ab} = g(e_a,e_b)$ on the bubble.

While this coordinate system is convenient to describe the geometry, due to the simplicity of the gluing conditions, the same results can be reproduced in another global coordinate system, denoted $(\tau, \rho, \phi)$, in which the $\ads_3^\pm$ metrics read
\begin{equation}\label{eq:other_coords}
ds_{\pm}^2 = L_{\pm}^2 \left( - \cosh^2\rho_\pm\, d\tau_\pm^2 + d\rho_\pm^2 + \sinh^2\rho_\pm \, d\phi_\pm^2 \right) \, .
\end{equation}
This turns the gluing condition into
\begin{equation}\label{eq:gluing}
L_- \sinh \rho_- = L_+ \sinh \rho_+ \, ,
\end{equation}
which induces a discontinuity $\rho$ that must be taken into account. There is also a corresponding discontinuity in $\tau$.

We note the $SO(2,2)$ isometry group of $\ads_3$ is broken by the above metric to the subgroup $SO(1,2)$ that keeps the nucleation event fixed, and under which the bubble wall and the two $\ads^\pm_3$ regions are all invariant.


\subsection{Thick Bubbles and Conformal Structure}

\begin{figure}[h]
    \centering
    \input{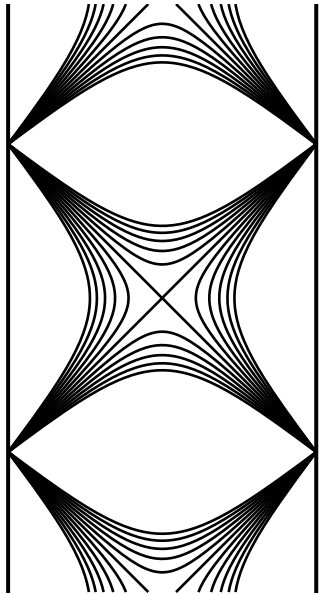}
    \caption{a cross-section of a Penrose diagram for $\ads$ spacetime with selected level sets of $\xi^2$, representing potential layers of a thick bubble. A choice of angular sector for $\tau$ eliminates the periodicity.}
    \label{fig:xipenrose}
\end{figure}

The metric described in the preceding section has a boundary with a ill-formed conformal structure, since the two semi-infinite cylinders corresponding to the conformal structures of the boundaries of $\ads^\pm_3$ are separated by a ring-like ``conformal singularity'', which builds up when the bubble reaches infinity. While this might seem an artefact of the thin-wall approximation, we have reasons to believe that this is not the case. In general, a ``thick'' bubble could be realised via a smooth metric with the same isometry group\footnote{Actually, the nucleation event cannot itself have such a symmetry which can only hold for sufficiently large bubble, well after nucleation. This is of course not an issue for what concerns the conformal structure of the boundary.} as a thin bubble, which is the $SO(1,2)$ subgroup of $SO(2,2)$ that keeps the nucleation centre fixed. Up to diffeomorphisms, the only invariant of this subgroup is 
\begin{equation}\label{eq:xi_invariant}
    \xi^2 \equiv \log\abs{\ch \rho \,\cos\tau} \, , 
\end{equation}
which generalises the flat-spacetime $r^2-t^2$, so that any candidate ``smoothed'' $L_\text{eff}$ can only depend on $\xi^2$ and, possibly, on a discrete choice of angular sectors\footnote{A single-bubble tunneling, for example, can be implemented letting $\Leff = L_-$ for $\tau <0$, a smooth function of $\xi^2$ for $0<\tau<\pi/2$, and $L_+$ for $\tau>\pi/2$.} for $\tau$. We have convinced ourselves that, independently of the smooth behaviour of the effective radius, the boundary value of $L_\text{eff}$ is still given by a step function, namely 
\begin{equation}\label{eq:L_eff_limit}
    \lim_{\rho\rightarrow \infty} L_\text{eff}(\tau,\rho) = \begin{cases}
    L_- & \tau < \frac{\pi}{2}\\
    L_+ & \tau > \frac{\pi}{2}
    \end{cases}\, .
\end{equation}
In geometric terms, all ``layers'' of the thick bubble can reach the boundary at the same time, and thus produce again a conformal singularity, separating two vacuum conformal structures. This is schematically depicted in the Penrose-like diagram of fig.~\ref{fig:xipenrose}. We remark that this structure is indeed imposed by symmetry, since it originates from a suitable Wick rotation of an $SO(3)$-invariant Euclidean solution. This is consistent with an intuitive picture in which each ``layer'' moves in a uniformly accelerated fashion, is asymptotically null and the slower ones start out closer to the boundary.
 
To conclude this section, let us briefly address the issue of gravitational collapse. It was shown~\cite{Coleman:1980aw,Abbott:1985kr} that $\ads$ thick bubbles nucleating inside Minkowski false vacua induce a ``big crunch'' due to a singular evolution of the scalar field $\Phi(\xi)$. However, the issue is subtler in the present case, since the proof in~\cite{Abbott:1985kr} rests on the existence of global Cauchy surfaces, which $\ads$ does not accommodate. To wit, the initial-value problem in global $\ads$ is ill-defined unless it is supplemented with appropriate boundary conditions.

However, the $SO(1,D-1)$ symmetry assumed in~\cite{Coleman:1980aw} and in the present work does not allow any plausible choice of boundary conditions. For instance, Dirichlet conditions for $\Phi(\xi)$ at the conformal singularity constrain it to be constant, since all slices of constant $\xi$ converge there\footnote{An analogous constraint holds for boundary conditions involving a finite number of derivatives.}. Regardless of how boundary conditions affect the issue at stake, we remark that the present work concerns primarily the expansion of the bubble, rather than the fate of $\ads_3^+$.

\section{The Holographic Entanglement Entropy}
\label{sec:holography}

In general terms, holographic dualities relate a gravitational theory to a non-gravitational one, typically a quantum field theory in a fixed background spacetime, in such a way that, whenever one side of the duality is strongly coupled, the other is weakly coupled and the two theories describe the same physics~\cite{Maldacena:1997re, Witten:1998qj, Gubser:1998bc, Aharony:1999ti}. The identification of the two theories then takes the form of a link between the bulk action and the boundary generating functional.

This prescription for holography has been employed to derive a number of important checks. Some of these have led to the Ryu-Takayanagi formula~\cite{Ryu:2006bv, Hubeny:2007xt, Rangamani:2016dms}, which relates entanglement entropy in the boundary theory and geometric quantities in the bulk, in a generalization of the Bekenstein-Hawking formula for black holes. In detail, the entanglement entropy of region $\mathscr{A}$ on the boundary is given by the extremal area of surfaces in the bulk whose boundary is $\partial \mathscr{A}$
\begin{equation}\label{eq:holo_EE}
S_{\text{ent}}(\mathscr{A}) = \inf_{\partial \mathcal{A} = \partial \mathscr{A}} \frac{\mathrm{Area}(\mathcal{A})}{4G_N} \, .
\end{equation}

The Ryu-Takayanagi formula is decorated by various corrections, arising for instance from higher curvature terms in the effective action for the bulk theory.

In light of its geometric simplicity, we take the Ryu-Takayanagi formula as a starting point and investigate the entanglement entropy functional of the boundary theory during the growth of the vacuum bubble. To this end, we essentially need to study the variational problem of finding the geodesic between two boundary points in the bubble geometry described in Section~\ref{sec:bubble_geometry}.

\subsection{Entanglement Entropy of the Bubble Geometry}
\label{sec:EE}

In accordance with the Ryu-Takayanagi formula, the entanglement entropy of a boundary interval $\mathscr{A} = A \bar{A}$ of size $2\theta_A$ follows from the (regularised) length of the shortest curve between its endpoints. The condition of extremality for a curve in the bubble geometry corresponds to it being composed, inside and outside the bubble, of segments of hyperbolic lines (of the relevant $\mathbb{H}^2$), joining with \textit{no kink} at the bubble wall.

\begin{figure}[h]
    \centering
    \scalebox{0.3}{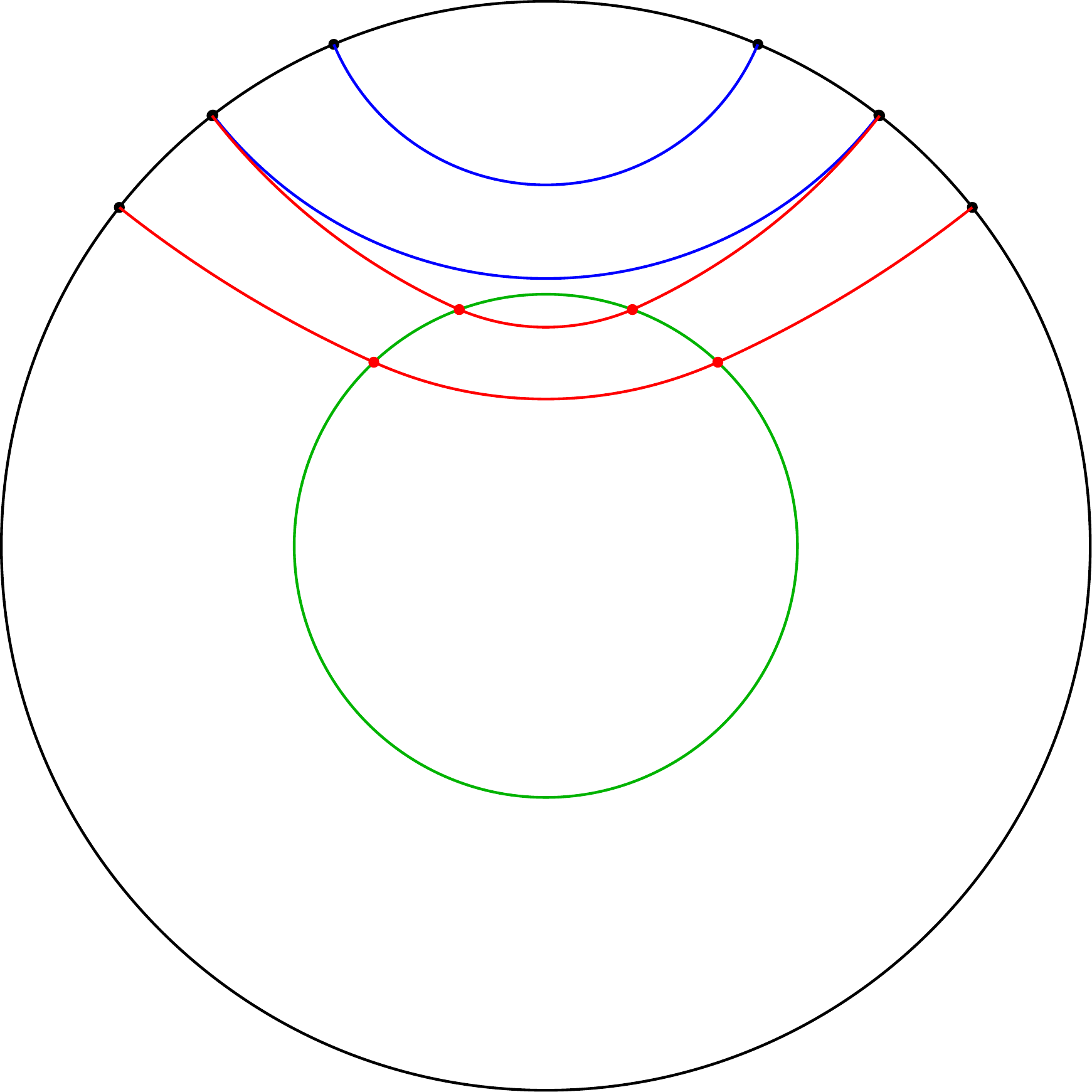}
    \caption{minimal curves for increasing $\theta_A$ in a double-Poincar\'e disk model. The two equal-length geodesics at the injection phase transition are depicted. Notably, the transition occurs before the vacuum geodesic becomes tangent to the bubble.}
    \label{fig:phasetransition}
\end{figure}

This no-kink condition is more precisely stated as the requirement that the slope $\dv{\ell}{\phi}$, where $d\ell \equiv (1+r^2/\Leff^2)^{-1/2}dr$ is the differential radial geodesic distance, be continuous across the bubble wall\footnote{The absence of a kink translates graphically into the condition that the geodesic segments be tangent in a conformal model, such as the ``double-Poincar\'e disk'' that we employ in fig.~\ref{fig:phasetransition}. Equivalently, the angles formed with a ray of the circle, measured in the inner and outer hyperbolic planes, coincide.}. It is explicitly a consequence of the (distributional) geodesic equation, which in the present case can be integrated to
\begin{equation}\label{eq:geodesic}
    \frac{dr}{ds} = \sqrt{\left(1+\frac{r^2}{\Leff^2}\right)\left(E- \frac{J^2}{r^2}\right) }\, ,
\end{equation}
where $E$ and $J$ are integration constants and $s$ an affine parameter, so that
\begin{equation}\label{eq:no_kink}
    \dv{\ell}{\phi} = \left(1+\frac{r^2}{\Leff^2}\right)^{-1/2} \, \frac{r^2}{J}\dv{r}{s}  = \frac{r^2}{J} \sqrt{E-\frac{J^2}{r^2}}
\end{equation}
is indeed continuous at the bubble wall. To explain it in a more intuitive fashion, ``zooming in'' on the intersection of the geodesic with the bubble and sending $L_\pm \rightarrow \infty$, one recovers the regular Euclidean plane, consistently with the absence of a kink.

We distinguish two possible phases for the extremal curve:

\begin{itemize}
    \item The \textbf{vacuum phase}, simply given by the hyperbolic line in $\mathbb{H}_-^2$ between two symmetric endpoints $A$ and $\bar{A}$, which only exists if $\cos \theta_A > \cos \theta_A^\text{par} \equiv  \tanh(r_\text{bubble}/L_- )$.
    \item The \textbf{injection phase}, where the curve injects into the bubble at a point $B$ at an angle $\theta_B$ from the interval centre, follows a line in $\mathbb{H}^2_+$ until the symmetric point $\bar{B}$, then exits the bubble and follows a line to $\bar{A}$. The angle $\theta_B$ is fixed by the no-kink condition.
\end{itemize}

In Appendix~\ref{sec:geodesicappendix} we derive both the no-kink condition, written as an equation suitable to numerics, and the length of the corresponding geodesics using hyperbolic geometry. Then, for each value of $\theta_A$ we first solve the no-kink equation for the injection phase numerically, and use the results to compare the lengths of the two phases to determine the minimal one.

\begin{figure}[h]
    \centering
    \scalebox{0.6}{\input{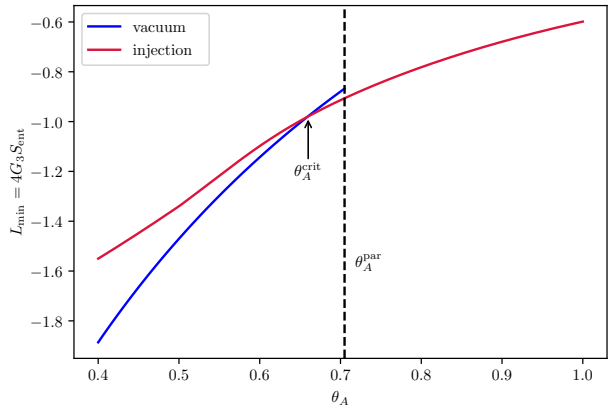}}
    \caption{finite part of geodesic length for the two phases plotted against boundary interval size. A cosmological constant ratio of $\frac{1}{2}$ has been chosen as an example.}
    \label{fig:phases}
\end{figure}

We find that the length of the injecting curve drops below that of the vacuum curve at a critical angle $\theta_A^\text{crit} < \theta_A^\text{par}$, marking a phase transition beyond which the penetrating geodesic is favoured.

\section{$c$-functions and the ``Bubble/RG'' Conjecture}\label{sec:RG}

In this section we introduce our picture of holographic vacuum decay via bubble nucleation. As previously mentioned, the entanglement structure induced by the bubble via the Ryu-Takayanagi prescription hints at some process which reduces the effective number of degrees of freedom on the boundary. Moreover, this process is necessarily irreversible, since bubble nucleation only occurs in the direction of decreasing vacuum energy. These features point to an holographic interpretation of non-perturbative\footnote{We emphasize that we require our original vacuum to be strictly metastable, namely stable against small fluctuations.} vacuum decay in terms of an RG flow. We can now follow a number of standard procedures to construct $c$-functions which appear to capture this type of scenario~\cite{Casini:2006es, Myers:2010xs, Myers:2010tj, Myers:2012ed, Albash:2011nq, Casini:2012ei}.

In the following we shall work in global coordinates, since Poincar\'e coordinates, which do not cover the whole of $\ads$, are problematic in the presence of a centred, axially symmetric bubble. The holographic RG framework is usually described in Poincar\'e coordinates, a feature which impacts the nature of the dual RG flow in the boundary theory in a non-trivial fashion. We shall return to this issue in more detail in Section~\ref{sec:horocyclic}, explaining how our framework incorporates the Poincar\'e holographic RG picture as a limiting case.

\subsection{$c$-functions from Entanglement Entropy}

As outlined in Section~\ref{sec:RG}, one can use the entanglement entropy to construct a $c$-function. Given a fixed spatial slice, taken out of the preferred foliation induced by the isometries of the bubble, the dependence of the entanglement entropy on the interval length $l$ will only feel the presence of the bubble for sufficiently large $l$, as explained in Section~\ref{sec:EE}. This, along with the fact that we are working in global coordinates where the conformal boundary of a spatial slice has the topology of a circle, suggests that $l$ is not the most relevant quantity to construct a $c$-function. Indeed, our aim is to relate the bubble expansion to an RG flow, and the interval length at \textit{fixed} time does not appear suitable in this respect, since a canonical definition of a boundary length scale at infinity appears problematic in global coordinates. This is to be contrasted with the Poincar\'e holographic RG, where intervals result from a stereographic projection onto the line and therefore the rescaling of interval lengths is reminiscent a coarse-graining procedure.

The relevant scales in the bulk are, instead, the coordinates $r \, , \eta$ which are related via eq.~\eqref{eq:bubble_motion}. This means that, at fixed time $\eta = \eta^*$, the bubble radius $R \equiv \eta^*$ appears as the only relevant scale from the boundary perspective, and motivates the choice of fixing an interval $\mathscr{A}$ of half-angle $\theta_A$, and of considering
\begin{equation}\label{eq:ent_c_fun}
c_{\mathscr{A}}(R) \equiv 3\,\theta\,\frac{d\sent(\theta;R)}{d\theta}\Big |_{\theta = \theta_A} \, .
\end{equation}
This provides an example of a $c$-function constructed out of the entanglement pattern of the system, although not necessarily the only one. The aforementioned identification of the bubble radius with an RG scale is the first step towards the proposed framework in which vacuum bubbles are associated with RG flows.

Furthermore, one can recast the dependence on $R$ in eq.~\eqref{eq:ent_c_fun} as a dependence on the interval half-angle $\theta$ in the following sense: instead of fixing $\mathscr{A}$, given the bubble radius $R$ one can take the critical interval size $\theta_A^\text{crit}$. This defines a correspondence $\theta(R)$ which may be employed to work with the angular size.

The most natural choices for $\mathscr{A}$ would be either half of the boundary, so that the corresponding entanglement entropy is immediately sensitive to the bubble upon nucleation in a smooth fashion, or the whole boundary\footnote{More precisely, one should take the limit as $\theta_A \to \pi/2^-$, since the full boundary has vanishing entanglement entropy.}, which interestingly gives a \textit{step function}: before the bubble arrives at the boundary $c_{\text{bdry}} = c_-$, while afterwards its value jumps to $c_{\text{bdry}} = c_+$, where $c_\pm$ are the central charges associated to $L_\pm$. The presence of the bubble does not influence the boundary until the very instant it touches it, at the end of the expansion. Notably, this happens in a finite coordinate time $t_\text{tot} = L_- \pi/2$ (or $\eta = \infty$) in the $\ads_3^-$ patch outside the bubble, which conceivably leaves open the possibility of multi-bubble events that could modify the boundary theory in different ways.

\begin{figure}
    \centering
    \input{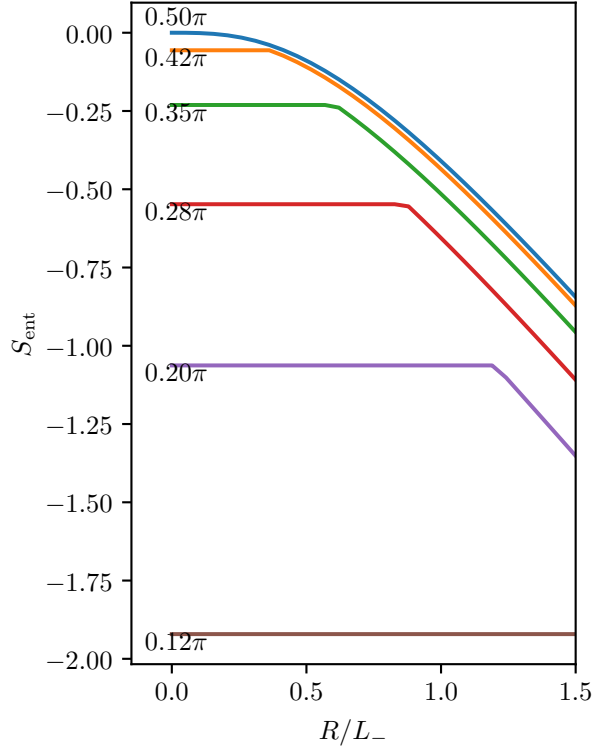}
    \caption{finite part of the entanglement entropy vs bubble radius, for various $\theta_A$. Notice the smooth behaviour of the $\theta_A = \pi/2$ curve, which corresponds to half of the boundary. This would translate into a smooth interpolating description for the dual RG flow.}
    \label{fig:EE_xm}
\end{figure}

\subsection{$c$-functions from the Null Energy Condition}

When looking for holographic $c$-functions, another option is applying the standard prescription~\cite{Anselmi:1997am, Girardello:1998pd, Henningson:1998gx, Freedman:1999gp, deBoer:1999tgo, Myers:2010xs, Myers:2010tj, Myers:2012ed} in global coordinates. This involves a procedure analogous to the one typically carried out in the Poincar\'e patch, which defines $c$-functions in terms of the exponential warp factors that appear in asymptotically $\ads$ metrics.

Indeed, our choice of writing the metric as in eq.~\eqref{eq:bubble_metric} conveniently defines a bulk $c$-function in terms of the effective radius $L_{\text{eff}}(\eta,r)$ given in eq.~\eqref{eq:eff_L}, extending the dictionary in eq.~\eqref{eq:c_dictionary}. An important difference with respect to the scenario outlined above, however, is that the resulting $c$-function is \textit{time-dependent}. Physically, this can be ascribed to the dynamical nature of the geometry, although the actual functional dependence can be recast in terms of the combination $r-\eta$ only. Indeed, once again one obtains a step function
\begin{equation}\label{eq:eff_c}
c_{\text{eff}}(\eta,r) \equiv \begin{cases}
c_-\, , & r < \eta\,,\\
c_+\, , & r > \eta\,,
\end{cases} \quad .
\end{equation}

One may readily observe that, when suitably extended beyond the collapse\footnote{This can be done, for instance, by gluing two coordinate charts, each of which would cover one of the $\ads_3^\pm$.}, these $c$-functions \textit{approach} $c_{\text{bdry}}$ as $r \to +\infty$, a reassuring consistency check, while their discontinuous nature can presumably be ascribed to the thin-wall approximation. The same cannot be said for the $c$-function defined by $c_{\text{bdry}}$, whose discontinuity is seemingly linked to the conformal singularity of the bubble geometry.

Notice that the monotonic behaviour of this $c$-function may appear compromised by the discontinuous nature of the geometry that we consider. However, the thin-wall regime is only an ideal limit of a smooth function, which interpolates between $L_\pm$, and hence between $c_\pm$. The monotonic behaviour of holographic $c$-functions reflects in general the null energy condition~\cite{Myers:2010xs, Myers:2010tj, Myers:2012ed}. As explained in Section~\ref{sec:horocyclic}, the computation can be reproduced for horocyclic bubbles, since it reduces to the case of a domain wall in Poincar\'e coordinates. A similar computation can be carried out in global coordinates, employing a smoothing of the singular metric~\eqref{eq:bubble_metric} of the form
\begin{equation}\label{eq:c-fun_metric}
    ds^2 = -\bigg(1 + \frac{r^2}{L^2} \bigg) \frac{d\eta^2}{\big(1+\frac{\eta^2}{L^2}\big)^2} + \frac{dr^2}{1 + \frac{r^2}{L^2}} + r^2 d\phi^2 \, ,
\end{equation}
where now $L(\eta,r)$ is a smooth function of $\eta$ and $r$. While this ansatz should have the correct form in the thin-wall regime, it would be interesting to investigate whether the exact Coleman-de Luccia instanton dictates a different one in the general case.

The two null energy condition (NEC) bounds
\begin{equation}\label{eq:nec_einstein}
    T_{\mu\nu} \, k_\pm^\mu \, k_\pm^\nu \geq 0 \, , \quad \text{with} \quad k_\pm = \frac{1+\frac{\eta^2}{L^2}}{\sqrt{1+\frac{r^2}{ L^2}}} \, \partial_\eta \pm \sqrt{1+\frac{r^2}{L^2}} \, \partial_r \, ,
\end{equation}
yield, using Einstein Equations 
\begin{equation}\label{eq:nec_bounds}
    \frac{\eta^2}{1+\frac{\eta^2}{L^2}} \, \pd_r L\, \geq \, \frac{r^2}{1+\frac{r^2}{ L^2}} \, \abs{\pd_\eta L} \, .
\end{equation}
These bounds further imply
\begin{equation}\label{eq:monotonicity}
    \pd_r L \, \geq \, 0 \, ,
\end{equation}
so that $r$ can be used as a holographic RG scale and
\begin{equation}\label{eq:nec_c-function}
c \equiv \frac{3L}{2G_3}
\end{equation}
is a $c$-function. Indeed, a constant $L$ saturates both NEC bounds.

\subsection{A Consistency Check: the Holographic Trace Anomaly}

As a final remark on other $c$-function constructions, and in order to provide further evidence for our proposal, let us briefly comment on an additional way to explore how the central charge of the boundary theory is affected by the bubble, the (holographic) trace anomaly. Two-dimensional quantum field theories on a curved spacetime with Ricci scalar $\mathcal{R}$ generally loose a classical conformal symmetry. In our case, this breaking reflects itself in an anomalous trace of the boundary energy-momentum tensor,
\begin{equation}\label{eq:trace_anomaly}
\langle {T^\mu}_\mu \rangle = - \, \frac{c}{12}\,\mathcal{R} \, .
\end{equation}

In Poincar\'e coordinates, the holographic computation of this anomaly has been carried out in~\cite{Henningson:1998gx}. Due to the dynamical nature of our problem, it is not clear \textit{a priori} whether the same procedure applies, but one expects that the time dependence should deform the anomaly in a manner compatible with replacing
\begin{equation}\label{eq:c_time_dep}
c \,\, \longrightarrow \,\, c_{\text{bdry}} \, .
\end{equation}
However, the standard prescription to compute the energy-momentum tensor VEV should still apply insofar as holography is valid, since we are assuming the Ryu-Takayanagi conjecture to begin with. While the computation, which still presents some subtleties, can be simplified focusing on the trace directly, we would like to stress that the traceless part of this VEV should provide quantitative information on how an \textit{off-centred} bubble affects the boundary theory. This issue will be the subject of a future investigation.

In computing the trace anomaly, one can attempt to generalize the procedure followed in~\cite{Balasubramanian:1999re}, whereby the boundary curvature in eq.~\eqref{eq:trace_anomaly} is recovered via the bulk extrinsic curvature. To this end, let us first emphasize that the general formula for the trace anomaly of the boundary theory,
\begin{equation}\label{eq:VEV_trace}
    \langle {T^\mu}_\mu \rangle = \,-\,\frac{1}{8\pi G}\,\big(\Theta + \Theta_{\text{c.t.}} \big) \, ,
\end{equation}
derived by the authors of~\cite{Balasubramanian:1999re}, will only hold in the present case if a term corresponding to the bubble energy-momentum tensor is added to the classical action. This is needed in order that the bubble geometry and scalar profile satisfy the bulk equations of motion, which also cancels the bulk contribution to the variation with respect to the boundary metric\footnote{Specifically, only its \textit{conformal class} matters.} $\gamma_{\mu \nu}$. Once this is done, it appears that the procedure can be extended to the case under consideration.

To begin with, one needs to modify the counterterm, which in $\ads_3$ is $2/L$. If $\langle T_{\mu \nu}\rangle$ is to be finite when evaluated on all classical solutions, we expect that a correct counterterm, which in any case has to reproduce $2/L_{\text{eff}}$ in the bubble geometry, should be expressed as a suitable function of the scalar potential. Then, writing a generic metric deformation in the form
\begin{equation}\label{eq:deformation}
    ds^2 = - \, f(\eta, r) \, \gamma_{\eta \eta}\,d\eta^2 + \frac{dr^2}{f(\eta, r)} + r^2\, \gamma_{\phi \phi} \,d\phi^2 + \frac{2r^2}{L_{\text{eff}}}\,\gamma_{\eta \phi}\, d\eta \,d\phi \, ,
\end{equation}
where $f(\eta,r) \equiv 1 + r^2/L^2_{\text{eff}}$, one can verify that it coincides with the one derived in Fefferman-Graham~\cite{Fefferman:1985, Fefferman:2007rka} coordinates\footnote{The conventions used in Section 3.2 of~\cite{Balasubramanian:1999re} rescale $\gamma$ by a factor $r^2$. In our convention, $\gamma$ has a finite boundary value.}, where the $\ads$ radius jumps from $L_-$ to $L_+$ after a finite time, provided one extends the coordinate system to include times after the bubble has reached the boundary. In fact, letting $n$ be the unit vector normal to the (regularized) boundary and $h_{\text{tr}}$ be the associated transverse metric, the bulk expression for the extrinsic curvature,
\begin{equation}\label{eq:invariant_extrinsic_curvature}
    \Theta = h_{\text{tr}}^{\mu \nu} \, \Gamma^A_{\mu \nu} \, n_A = - \frac{1}{2} \, \sqrt{f(\eta, r)} \, g^{\mu \nu} \, \partial_r \, g_{\mu \nu} \, ,
\end{equation}
gives the same result when evaluated in a Fefferman-Graham patch, since depending on whether the bubble has arrived at the boundary $f(\eta, r) \sim r^2/L^2_{\pm}$.

This also shows that the boundary deformation $\gamma$ is the correct counterpart of the Fefferman-Graham one, as one may observe from the large-$r$ asymptotics. Indeed, going back from $\eta$ to the standard global time coordinate $t$, the transformed $\gamma_{tt}, \gamma_{t\phi}$ comprise, alongside $\gamma_{\phi \phi}$, the deformation parameters which correspond to the (conformal class of the) boundary metric
\begin{equation}\label{eq:boundary_deformation}
    ds^2_{\text{bdry}} = -\,\gamma_{tt}\,dt^2 + \gamma_{\phi \phi}\,L^2_{\pm} \,d\phi^2 + 2L_{\pm} \, \gamma_{t \phi}\,dt\, d\phi \, ,
\end{equation}
which dominates in eq.~\eqref{eq:deformation} for large $r$, since
\begin{equation}\label{eq:asymptotics_boundary_deformation}
    ds^2 \sim \frac{r^2}{L^2_{\pm}} \, ds^2_{\text{bdry}} \, ,
\end{equation}
again depending on whether the bubble has arrived at the boundary.

Furthermore, one may verify that any smooth deviation from $L_{\text{eff}}$, which can also be defined for a thick bubble, does not contribute to the boundary asymptotics, consistently with the fact that, even outside the thin-wall approximation, the conformal structure of the boundary presents a singularity. To put it more simply, the boundary always sees the whole bubble arriving at the same instant. Hence, the trace anomaly
\begin{equation}\label{eq:trace_anomaly_result}
\langle {T^\mu}_\mu \rangle = - \, \frac{c_{\text{bdry}}}{12}\,\mathcal{R}
\end{equation}
indeed obtains with the replacement of eq.~\eqref{eq:c_time_dep} and the counterterm $\Theta_{\text{c.t.}} = 2/L_{\text{eff}}$.

In summary, this analysis shows that the deformation $\gamma_{\mu \nu}$ correctly corresponds to the Fefferman-Graham one, and the expectation of a step-like $c$-function from the trace anomaly is reproduced, alongside the absence of contributions due to deviations from a thin bubble. In addition, the framework we employed can be readily extended to generic (multi-)bubble configurations. Thus one may conclude that, in some sense, the holographic entanglement entropy provides a better probe of the physics, since it can detect the bubble arrival in a smooth fashion.


\section{Integral Geometry and Off-centred Bubbles}

A natural question concerns the holographic interpretation of the site of the nucleation event and, in particular, the modification of the RG flow for off-centred bubbles. For the purpose of performing $SO(1,2)$ hyperbolic translations to investigate this issue, we find it convenient to reformulate the correspondence in the formalism of Holographic Integral Geometry~\cite{Czech2015}\footnote{For an earlier work on RG flows and Integral Geometry, albeit in a different setting, see~\cite{Bhowmick:2017egz}.}, which we review in Appendix~\ref{sec:igappendix}.

Specifically, we consider the Crofton form $\omega/4G_3$ instead of $S_\text{ent}$, since the former is insensitive to the cutoff and contains no divergent part. The reason is that an $SO(1,2)$ isometry to translate the bubble behaves unwieldily in the presence of divergent terms: it deforms the cutoff surface which has then to be brought back to its original position. Equivalently, the finite part of $S_\text{ent}$ is not an $SO(1,2)$ scalar because the extraction of the finite part is not invariant. $\omega$ is instead a finite and covariant two-form. In particular, the ratio\footnote{This is possible only because $\omega$ is a form of top rank in the present case.} to the vacuum $\ds_2^-$ volume form, defined by
\begin{equation}\label{eq:Omega_ratio}
\omega(u,v) = \Omega(u,v) \,\vol_{\ds_2^-} \, ,
\end{equation}
is a finite scalar field on $\mathcal{K}_2$. Therefore, one may exploit this fact to study off-centred bubbles applying $\sltr \rightarrow SO(1,2)$ transformations to data and conclusions already obtained in the central case. These transformations appear in the triplicate role of asymptotic bulk isometries, kinematic symplectomorphisms, and boundary restricted conformal maps.

\begin{figure}
    \centering
    \scalebox{0.8}{\input{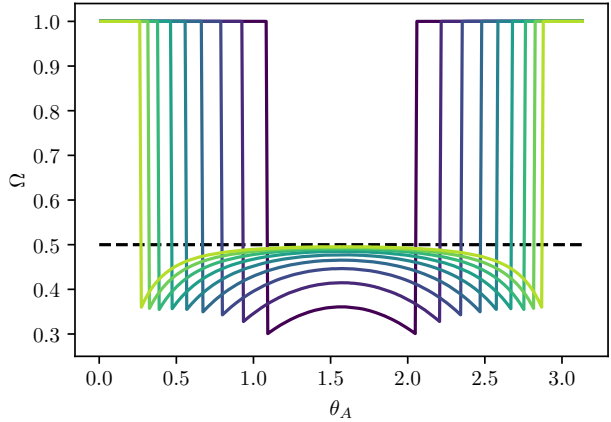}}
    \caption{the relative Crofton factor $\Omega$ for a centred bubble as a function of interval size, for increasing values of the bubble radius (dark to light). A cosmological constant ratio of $\frac{1}{2}$ has been chosen as an example, which leads to the limiting values $\Omega_- = 1$, $\Omega_+ = \frac{1}{2}$. The $\delta$-function wall at the phase transition is not depicted.}
    \label{fig:croftonvacuum}
\end{figure}

For centred bubbles,  $\Omega$ only depends on the combination $\theta_A = \frac{u-v}{2}$, the boundary interval half-size\footnote{In the language of $\ds_2$, $\theta_A$ is diffeomorphic to $\ds_2$ time in closed slicing, and $\phi$ is the coordinate on the slice.}, and not on the boundary centre $\phi = \frac{u+v}{2}$. $\Omega(\theta_A)$ displays an external $\delta$-function wall corresponding to the injection phase transition described in Section~\ref{sec:EE}. \textit{Outside} the wall $\Omega=1$, the constant value related to the original vacuum, while \textit{inside} the wall one finds a smooth dependence approaching the constant value associated to the new vacuum, which is the ratio of the cosmological constants (see fig.~\ref{fig:croftonvacuum}).

Shifting the bubble corresponds to a boost in $\ds_2$, which induces a mixing between the $\theta_A$ and $\phi$ coordinates or, more suggestively, between boundary momenta and positions, a feature which we shall discuss in Section~\ref{sec:generalisedrg}. The $\delta$-function wall in $\Omega$ is deformed into an ellipse in $\ds_2$. Intuitively, when the bubble is off-centred, boundary intervals closer to it will begin to be affected at smaller sizes (see fig.~\ref{fig:kineplots}). Hence, the deformed entanglement pattern on the boundary should encode this effect with some spatial localisation, and should evolve under the flow in a manner reminiscent of the corresponding bulk bubble expansion.

\section{Off-centred Renormalisation}
\label{sec:generalisedrg}

When the symmetries of the decay geometry are taken into account, it becomes impossible to match the growth of a centred bubble with a standard holographic renormalisation procedure, which is normally implemented as a sequence of decimations and rescalings within a Poincar\'e chart~\cite{Myers:2010xs, Myers:2010tj, Myers:2012ed}. Poincar\'e rescalings do not map to an isometry of a centred bubble, which instead has an $SO(2)$ subgroup of rotational isometries. We propose that, instead, the precise prescription for a centred bubble is a renormalisation procedure that respects this rotational symmetry, and is schematically implemented as a decimation and rescaling of the angular $\phi$ coordinate. Since the radius of the boundary circle shrinks under such an RG flow, and would naively vanish in an infinite RG time, this ought to be counteracted by a preemptive blowup of the circle in the original, undeformed $\cft_-$. As a result, one should explore simultaneous limits of initial blowup and total RG flow time. We conjecture that theories with a holographic bulk dual do not degenerate under this limit and approach a non-trivial infrared $\cft_+$, which would reflect the existence of a stable final $\ads_+$ classical vacuum in the bulk.

In addition, if one imagines to extend the proposed ``bubbleography'' correspondence to cases in which tunneling to \textit{bubbles of nothing}~\cite{Witten:1981gj} can occur, the preceding discussion implies that such scenarios would conceivably lead to trivial endpoints of the dual RG flow: in this context, the expansion would leave behind an $\ads$ geometry of vanishing radius. For previous discussions on the holographic interpretation of bubbles of nothing, see~\cite{Balasubramanian:2002am, Balasubramanian:2005bg, He:2007ji}.

In any case, a central renormalisation procedure respecting the rotational symmetry would allow to define a renormalisation step for off-centred bubbles simply as the central RG step conjugated by the $SO(1,2)$ isometry that shifts the bubble. Analogously, bubble nucleation should again correspond to a relevant deformation, up to the same $SO(1,2)$ conjugation. Equivalently, there is a boundary picture in which the deformation is space-dependent\footnote{A simpler instance can be realised in a theory with a space-dependent running cutoff scale $\Lambda(x)$.}, and the RG flow proceeds also partially in position space.

\begin{figure}[h]
    \centering
    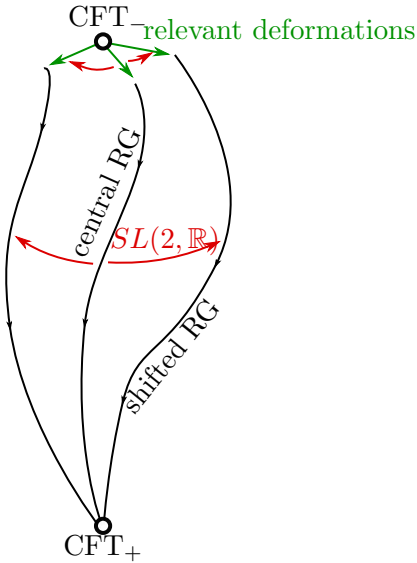
    \caption{a schematic depiction of the family of generalised relevant deformations followed by generalised RG flows, all connected by $\sltr$ transformations.}
    \label{fig:rgflower}
\end{figure}

\subsection{Recovering a Poincar\'e RG}\label{sec:horocyclic}

\begin{figure}
    \centering
    \input{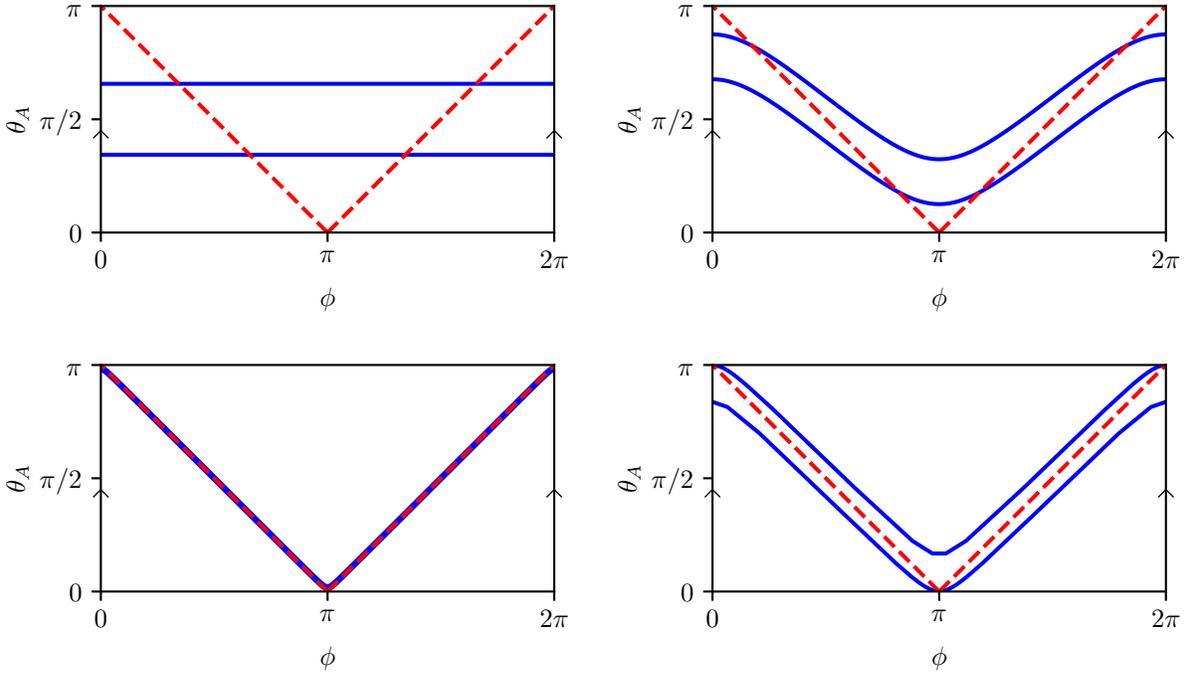}
    \caption{the $\delta$-function wall in the Crofton factor $\Omega$, the locus of the injection phase transition (blue), depicted in $\mathcal{K}_2$ using the $(\theta_A,\phi)$ chart, which is conformal for $\ds_2$. In all cases $\rho_- = 0.5$, $L_-/L_+ = 0.5$. Upper left: for a centred bubble. Upper right: after an $SO(1,2)$ boost with $\beta=0.8$. We remark that this introduces a $\phi$-dependence. Bottom left: after a $\beta=0.999$ boost, the walls converge to the marked lightcone (red). Bottom right: again $\beta=0.999$, but with a suitable rescaling of $\rho_-$.}
    \label{fig:kineplots}
\end{figure}

A bubble translated infinitely far away from the origin, with its radius $R$ suitably rescaled in such a way that the wall remains at a finite distance from the origin\footnote{This limit is certainly sensible, since for bubbles in $\ads$ the radius $R$ becomes infinite in finite coordinate time $t$, and can thus be made arbitrarily large with a small time translation. This is depicted in the bottom-right numerical plot in fig.~\ref{fig:kineplots}.}, is a particular limiting case. This is actually, in a sense, the most likely scenario, since tunneling is favoured by the exponentially large bulk volume fraction that lies far away from the origin for large cutoff. In the limit, the bubble wall becomes a travelling horocycle, and the corresponding dual RG flow is simply the standard holographic RG procedure in Poincar\'e coordinates. Indeed, the horocyclic bubble at each time is precisely a $z=\mathrm{const}.$ curve in a Poincar\'e chart.

In addition, one may conceive multiple bubble nucleations occurring within the time frame of a single expansion. This should allow for the construction of a larger and diverse family of deformations and RG flows from $\cft_-$ to $\cft_+$, since the characteristic step-like behaviour of $c$-functions provides a natural building block for a variety of scenarios.

\subsection{The Dual Relevant Deformations}

The identification of the relevant deformation of the original $\cft_-$ corresponding to the nucleation event remains an important open problem. Explicit top-down realisations of the scenario discussed here should be relevant in order to address it, since they typically bring along a more transparent description of the corresponding holographic duals. This could also provide an additional handle to perform more in-depth analyses of the RG flow studying, for instance, the scaling of correlation functions.

In principle, one may expect that such a relevant deformation could be related to the decay width (per unit volume) associated to the tunneling process, which may be computed via standard techniques in the semi-classical limit~\cite{Coleman:1977py, Callan:1977pt, Coleman:1980aw}. Indeed, in the classical limit tunneling is completely suppressed, and the starting point of the flow ought to approach the original $\cft_-$, which remains fixed.


\section{Conclusions}

In this paper we outlined a connection between non-perturbative vacuum decay in $\ads$ Quantum Gravity and RG flows between (deformed) CFTs. We studied the holographic properties of a vacuum bubble in $\ads_3$, which nucleates and expands via semi-classical tunneling, focusing on the entanglement entropy and $c$-functions. The former appears to capture (not necessarily universal) properties of the intermediate flow in a smooth fashion.
Our analysis confirms that a connection exists between vacuum decay processes and RG flows, which we intend to develop further in a future work. The formalism of Integral Geometry captures conveniently and elegantly the relevant physics, and we are investigating its role in more general scenarios. The study of correlation functions and of the energy-momentum tensor of off-centred and multi-bubble configurations could provide further insights, while explicit top-down constructions involving non-supersymmetric string theories~\cite{AlvarezGaume:1986jb, Sagnotti:1995ga, Sagnotti:1996qj}, such as those arising from ``Brane Supersymmetry Breaking''~\cite{Sugimoto:1999tx, Antoniadis:1999xk, Angelantonj:1999jh, Aldazabal:1999jr, Angelantonj:1999ms, Dudas:2000nv, Pradisi:2001yv, Angelantonj:2002ct, Mourad:2017rrl}, or orbifolds of supersymmetric ones~\cite{Kachru:1998ys,Lawrence:1998ja,Bershadsky:1998mb,Bershadsky:1998cb,Schmaltz:1998bg,Erlich:1998gb}, might also provide a fruitful avenue of investigation. For instance, the $\mathbb{Z}_k$ orbifold of~\cite{Horowitz:2007pr} features a bubble of nothing and a known holographic dual~\cite{Harlow:2010az}, which is a quiver-like non-supersymmetric $U(N)^k$ gauge theory.

To conclude, we can comment on some potential implications. At present, vacuum stability in Quantum Gravity poses significant theoretical challenges, even at the semi-classical level. Hence, classifying criteria for stability appears of primary importance, and some properties that stable (classical) vacua should possess have already surfaced, a prime example being the \textit{Weak Gravity Conjecture} (WGC)~\cite{ArkaniHamed:2006dz}.
As explained in~\cite{Freivogel:2016qwc, Ooguri:2016pdq, Danielsson:2016rmq}, it appears that if the WGC (or a similar stability criterion) holds, nucleation events should continue to occur at least until a supersymmetric $\ads$ classical vacuum is reached. This is because, in the supersymmetric case, stability prevents tunneling, and only domain walls can be present~\cite{Cvetic:1992st,Ceresole:2006iq, Cvetic:1996vr, Bandos:2018gjp}. In the RG picture that we presented the IR stable endpoint of the flow would then be supersymmetric, which resonates with the phenomenon of {\it emergent supersymmetry} in some condensed matter systems\footnote{See e.g.~\cite{Friedan:1984rv} or, for a modern review,~\cite{Lee:2010fy}, and references therein.}.
It would be interesting to explore whether the framework that we described can be used as a tool to address vacuum stability in more intricate contexts from the perspective of better-understood RG flows, which can then be approached with powerful analytic and numerical techniques.

\section*{Acknowledgements}
We would like to thank S. Giusto, L. Martucci and D. Seminara  for useful discussions. We would especially like to thank A. Zaffaroni for valuable correspondence and for having brought to our attentions some references after reading the manuscript.

We are deeply grateful to A. Sagnotti for assistance and support to all authors during the realization of this project, and for his feedback on the manuscript. AB would like to thank N. Cribiori and A. Galliani for useful discussions, and especially  S. Lanza for having brought this topic to his attention. 

We also wish thank the Galileo
Galilei Institute for Theoretical Physics (GGI) for its hospitality during the ``LACES 2017'' (Lezioni Avanzate di Campi e Stringhe) Ph. D. School. AB wishes thank the GGI for its hospitality during the conference ``50  Years of the Veneziano Model: From Dual Models to Strings, M-Theory and beyond''.  

AB is supported by the Padova University Project CPDA119349 and by INFN-GSS. RA and IB are supported in part by Scuola Normale Superiore and by INFN (IS CSN4-GSS-PI).


\appendix

\section{Derivation of the Minimal Geodesic}\label{sec:geodesicappendix}

\subsection{The No-kink Condition}\label{sec:nokinkappendix}

\newcommand{\alphaout}{\alpha_\text{out}}
\newcommand{\alphain}{\alpha_\text{in}}

\begin{figure}
    \centering
    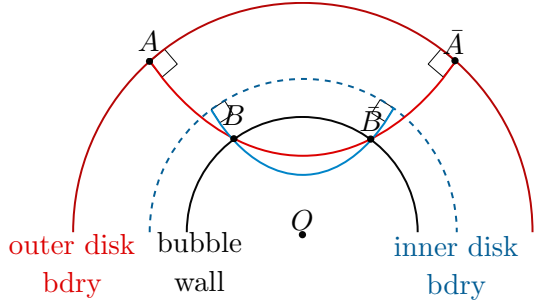
    \caption{the double-Poincar\'e model with a candidate injection polygonal.}
    \label{fig:doublepoinc}
\end{figure}

In order to give a visual representation of the geometry of a constant-time slice in the presence of the bubble, which consists of two hyperbolic planes $\mathbb{H}^2_\pm$ of different radii suitably glued along a circle, we employ a conformal model constructed from two superimposed and glued Poincar\'e disks, relatively scaled in such a way that the glued circles have the same size, as in fig.~\ref{fig:doublepoinc}. In the same figure, we marked a candidate polygonal for the injection-phase geodesic between boundary points $A$ and $\bar{A}$. To find an actual geodesic it is necessary to determine the point $B$ such that the no-kink condition is satisfied and, since the model is conformal, the kink also disappears graphically. Rotating the model such that there is symmetry about the vertical axis, we define $\theta_A$ and $\theta_B$ as the angles that the segments $OA$ and $OB$, respectively, make with the vertical. One may observe that $2\theta_A$ is the size of the boundary interval, which is a given parameter.

The no-kink condition is then equivalent to the statement that the angles that the hyperbolic line segments $AB$ and $B\bar{B}$ make with the bubble radius through $B$, which we name $\alphaout$, $\alphain$ respectively, be equal.

Let us now consider the inner disk. Let $C$ be the intersection between $B\bar{B}$ and the radius that bisects the $BO\bar{B}$ angle or, equivalently, $AO\bar{A}$, as one may observe via a symmetry argument. Since $\theta_B = BOC$, noting that $\alphain = CBO$ and that $OCB$ is right one finds, from the trigonometry of hyperbolic right triangles, that
\begin{equation}\label{eq:alphaouteqt}
    \cosh \rho_+ = \cot \alphain \, \cot \theta_B \, ,
\end{equation}
where $\rho_+$ is the geodesic radius of the bubble divided by $L_+$. Equivalently, the circumference of the bubble is $2\pi L_+ \sinh \rho_+$.

\begin{figure}
    \centering
    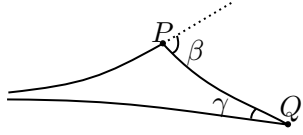
    \caption{an obtuse omega triangle.}
    \label{fig:omega}
\end{figure}

For what concerns the outer disk, let us first show an identity for omega triangles, namely hyperbolic triangles with exactly one ideal vertex. If an obtuse omega triangle has angles marked as in fig.~\ref{fig:omega}, then the length of segment $PQ$ is given by 
\begin{equation}\label{eq:omegatriang}
    PQ/L = \cosh^{-1}{\csc{\gamma}} - \cosh^{-1}{\csc{\beta}} \, ,
\end{equation}
where $L$ is the radius of the hyperbolic plane. This readily follows from dropping the perpendicular from $Q$ to the opposite side and making use of the formula for the angle of parallelism.

Let us now turn to the entire outer $\mathbb{H}^2_-$, including the portion that has to be excised for gluing, and consider the obtuse omega triangle $OBA$ in this plane. One may obseve that the obtuse angle $OBA$ is supplementary to $\alphaout$, and that $AOB = \theta_A - \theta_B$. Therefore, using eq.~\eqref{eq:omegatriang} one finds
\begin{equation}\label{eq:alphaineqt}
    \rho_- = \cosh^{-1}\csc(\theta_A-\theta_B) - \cosh^{-1} \csc\alphaout
\end{equation}
where, as in the preceding case, $\rho_- L_-$ is the geodesic radius of the bubble, but now measured in the \textit{original} $\mathbb{H}^2_-$ as if the interior $\mathbb{H}^2_+$ region were not present. More specifically, $\rho_\pm$ are related by the gluing condition in eq.~\eqref{eq:gluing} and are therefore not independent. Indeed,
\begin{equation}\label{eq:gluing_condition_radius}
L_+ \sinh \rho_+ = L_- \sinh \rho_- = r \, .
\end{equation}

Imposing the no-kink condition $\alphain = \alphaout$ from eqs.~\eqref{eq:alphaineqt} and~\eqref{eq:alphaouteqt} then yields the following transcendental equation for $\theta_B$,
\begin{equation}
    \sqrt{1+(\cosh\rho_+ \tan\theta_B)^2} = 
    \cosh(\cosh^{-1} \csc(\theta_A-\theta_B)-\rho_-) \, ,
\end{equation}
which we solve numerically, alongside the constraint $\abs{\alpha_{\text{in},\text{out}}}< \frac{\pi}{2}$. We find that there is exactly one solution for $\theta_B$ in this range for all values of the parametres.

\subsection{The Geodesic Length}

In order to compute the actual length of the geodesic, it is convenient to employ to an hyperboloid model in place of the disk model. Namely, we embed $\mathbb{H}^2_\pm$ as the locus $\{X^\mu X_\mu = -1 \, , \, X^0 > 0\}$ in $\mathbb{R}^{1,2}$, so that the geodesic distance between two points $P$ and $Q$, in terms of their embedded images $P^\mu$, $Q^\mu$, is given by
\begin{equation}\label{eq:hyperboloidmodel}
    d(P,Q) = L_\pm \cosh^{-1}(P^\mu Q_\mu) \, .
\end{equation}
Since the length is divergent for points on the boundary, we regularise $A$ by placing it on a cutoff surface at a large, but finite, geodesic distance\footnote{One may observe that $\Lambda$ is exponential in a cutoff on the global coordinate $r$. It can be identified with the UV cutoff usually employed in holography.} $\Lambda$ from the origin of $\mathbb{H}^2_-$. The length of the segments $AB$, $B\bar{B}$ and $\bar{B}\bar{A}$ can be computed via eq.~\eqref{eq:hyperboloidmodel}, and the resulting total, which determines the entanglement entropy, is
\begin{equation}
\begin{aligned}\label{eq:geolength}
    \mathcal{L}
={} & 2 L_- \Lambda \\  +{} & 2L_- \log[ \cosh \rho_- - \sinh \rho_- \cos(\theta_B-\theta_A) ]  + L_+ \cosh^{-1}[\cosh^2 \rho_+ - \sinh^2 \rho_+ \cos(2\theta_B)] \\ +{} & \mathcal{O}(\Lambda^{-1}) \, .
\end{aligned}    
\end{equation}
Once $\theta_B$ has been determined from the no-kink condition, one may insert it into eq.~\eqref{eq:geolength} to obtain a numerical estimate of the finite part of the length, and thus of the entanglement entropy according to the Ryu-Takayanagi formula.

\section{A Brief Review of Holographic Integral Geometry} \label{sec:igappendix}

In this section we review the basics of Integral Geometry in the hyperbolic plane, since it concerns the specific case of $\ads_3/\cft_2$. A more comprehensive review can be found in~\cite{Czech2015}.

In this context the main object of interest is the topological space of ``lines'' in an asymptotically $\mathbb{H}^2$ bulk time-slice, namely the set of extremal curves between two boundary points, which constitutes the kinematic space $\mathcal{K}_2$. This is a two-dimensional surface that has a natural symplectic (or equivalently Lorentzian) structure, the \textit{Crofton form}, induced from the (finite part of the) length $\mathcal{L}$ of curves in $\mathcal{K}_2$ via
\begin{equation}
     \omega(u,v) \equiv \pdv{\mathcal{L}(u,v)}{u}{v} \, du \wedge dv  = 4G_3 \pdv{S_\text{ent}(u,v)}{u}{v} \, du \wedge dv \, ,
\end{equation}
where $G_3$ is the three-dimensional Newton constant and $u$, $v$ are angular coordinates for the endpoints on the $\mathbb{S}^1$ boundary. The last equality holds assuming the Ryu-Takayanagi formula, and the Crofton form $\omega$ also affords an information-theoretic interpretation in terms of mutual conditional information~\cite{Czech2015}. In addition, one may define an induced Lorentzian metric
\begin{equation}\label{eq:kinetic_metric}
    ds_{\mathcal{K}_2}^2 \equiv \pdv{\mathcal{L}}{u}{v} \, du \, dv \, .
\end{equation}

In the case of an $\ads_3$ vacuum, indeed composed of $\mathbb{H}^2$ slices, the Crofton form reduces to
\begin{equation}
    \omega_0(u,v) = \frac{L}{2\sin^2(\frac{u-v}{2})} \, du \wedge dv \, ,
\end{equation}
which is actually the only $SO(1,2)$-invariant $2$-form on kinematic space up to rescalings. Indeed, $\omega_0 = \vol_{\ds_2}$ is the volume form on two-dimensional de Sitter spacetime\footnote{Some intuition on this obtains embedding $\mathbb{H}^2$ in $\mathbb{R}^{1,2}$ as a two-sheeted hyperboloid, where hyperbolic lines arise from intersection with time-like planes through the origin. Such planes are in a one-to-one correspondence with their unit space-like normal vectors, which lie in the $\ds_2$ one-sheeted hyperboloid.}, and $\mathcal{K}^{(0)}_{2}$ is naturally endowed with the Lorentzian structure of $\ds_2$.

For a general deformed metric, which is still asymptotically $\mathbb{H}_-^2$ in any constant-time slice, one finds that the corresponding kinematic space is $\ds_2^-$ asymptotically, in the limit of large (absolute) de Sitter time, while the central region of small $\abs{t}$ is modified. In the kinematic picture, which acts as an intermediary, the vacuum bubble originates as a perturbation in $\omega$ from $\vol_{\ds_2}$, which is localised around the throat and expands symmetrically in the $\ds_2$ past and future as bulk time progresses, establishing in its interior the \textit{new} $\ds_2^+$, of different radius, associated with $\mathbb{H}^2_+$.

The relevance of this construction comes from a classical theorem of Crofton, which states that the length of any (not necessarily geodesic) bulk curve $\gamma$ can be computed in terms of an area in $\mathcal{K}_2$, namely
\begin{equation}\label{eq:crotfon_thm}
    \mathcal{L}[\gamma] = \frac{1}{4} \int_{\kappa \in \mathcal{K}_2} \omega(\kappa)\, n_{\gamma,\kappa}
\end{equation}
where $n_{\gamma,\kappa}$ is the (signed) intersection number of the curves $\gamma$ and $\kappa$. Hence, excluding shadow effects~\cite{Hubeny:2013gta, Freivogel:2014lja}, which are absent in this case, the bulk geometry is completely reconstructable from the Crofton form, which therefore provides an amount of information equivalent to the full entanglement entropy.



\bibliographystyle{apsrev4-1}
%

\end{document}